\documentclass[a4paper,12pt,oneside,dvips,driver]{article}
\usepackage[english]{babel}
\usepackage{pstricks,pst-node,pst-coil}
\usepackage{epsfig}
\usepackage{amscd}
\usepackage{amssymb}
\usepackage[T1]{fontenc}
\usepackage{latexsym}
\usepackage[]{psfrag} 

\psfrag{el}{$\tilde{l}$}
\psfrag{eta0}{$\eta/(\sigma\xi_B)$}
\psfrag{tau1}{$\tau_1/\sigma\;$}
\psfrag{ggggtau2}{$\;\tau_2/\sigma$}
\psfrag{ft1t2ggh}{$\sigma f(\tau_1,\tau_2)\qquad $}

\newcommand{\beq}{\begin{equation}}
\newcommand{\eeq}{\end{equation}}
\newcommand{\bdm}{\begin{displaymath}}
\newcommand{\edm}{\end{displaymath}}

\begin{document}

\author{P. Jakubczyk, M. Napi\'{o}rkowski  \\
Instytut Fizyki Teoretycznej, Uniwersytet Warszawski \\ 00-681 Warszawa, Ho\.za  69, Poland \\
and \\
A. O. Parry \\
Dept. Mathematics, Imperial College \\
180 Queen's gate, London SW7 2BZ
} 
\title{Point tension in adsorption at a chemically inhomogeneous substrate in two dimensions} 
\date{} 
\maketitle 

\begin{abstract}
{We study adsorption of liquid at a one-dimensional substrate composed of a single chemical inhomogeneity of width $2L$ placed on an otherwise homogeneous, planar, solid surface. The excess point free energy $\eta (L,T)$ associated with the adsorbed layer's inhomogeneity induced by the substrate's chemical structure is calculated within exact continuum transfer-matrix approach. It is shown that the way $\eta (L,T)$  varies with $L$ depends sensitively on the temperature regime. It exhibits logarithmic divergence as a function of $L$ in the limit $L\to\infty$ for temperatures such that the chemical inhomogeneity is completely wetted by the liquid. In the opposite case $\eta (L,T)$ converges for large $L$ to $2\eta_0$, where $\eta_0$ is the corresponding point tension, and the dominant $L$-dependent correction to $2\eta_0$ decays exponentially. The interaction between the liquid layer inhomogeneities at $-L$ and $L$ for the two temperature regimes is discussed and compared to earlier mean-field theory predictions.} \\  


\noindent{\noindent}
\end{abstract}

\newpage

\section{Introduction}
\renewcommand{\theequation}{1.\arabic{equation}} 
\setcounter{equation}{0}
\vspace*{0.5cm}
Adsorption phenomena taking place at substrates equipped with geometrical and chemical structures have been attracting considerable interest in recent years [1-17]. It is stimulated by the development of experimental techniques allowing one to imprint substrates with patterns down to nanometer scale (see e.g.[3-6]), and a number of theoretical results predicting interesting adsorption behaviour for systems with broken translational invariance along the substrate. These include filling transitions [7-10], discontinuous changes of equilibrium droplet shapes as function of its volume [11-12], or condensation-type phase transitions [13]. 
In case of convex substrates, like spheres or cylinders, the wetting transitions are altogether excluded by the substrate geometry [14-17].  

The basic theoretical tool to describe adsorption phenomena at microscopic level is the well-developed density functional theory \cite{EvansDFT}. However, in numerous applications it suffices to apply a reduced, coarse-grained description based on effective Hamiltonian models. This approach captures the system's essential features close to surface phase transitions and often allows insight into fluctuation effects, which play an important role in the case of systems with short-ranged intermolecular interactions (e.g. in the case of 3d wetting). As was demonstrated in Refs [19-21], thermal fluctuations often modify the critical singularities of line tension associated to three-phase contact lines (e.g. affect the values of the corresponding critical indices), and strongly influence the properties of the contact line itself [21], especially in two-dimensional systems.  
 
Adsorption at planar substrates equipped with a single, well localized chemical inhomogeneity has been studied both from the point of view of the adsorption layer morphology and the properties of linear excess free energy [22-26]. However, most of these studies referred to mean-field (MFT) approximation, within which fluctuations of the interfacial shapes are neglected. An exception is Ref.\cite{Conf} where the droplet height and height-height interfacial correlations were studied in the asymptotic regime close to the inhomogeneity center, far away from the boundaries. The motivation for theoretical investigations of this kind of systems is provided by experimental works (see e.g. Refs [3-5]), in which solid substrates are imprinted with chemical patterns and brought into contact with fluid. Structures of the adsorbed liquid-like layers are then investigated as function of thermodynamic parameters and the pattern's characteristics. In recent experiments the chemical patterns extensions have approached nanometer scale, at which effective interactions between the fluid heterogeneities induced by the chemical structure might become pronounced. An expression describing the aforementioned effective interactions has been derived in Ref.\cite{P1} within mean-field approach; it will be briefly recalled in Section 4. The influence of interfacial fluctuations on these interactions and the associated linear contributions to the system's free energy has not been discussed so far.

In this work we aim at filling this gap by considering adsorption in a two-dimensional system whose partition function can be evaluated exactly. By decomposing the system's free energy into bulk, line, and point terms, one obtains the excess point contribution, which is analyzed as function of the system parameters. We show, that the system's behaviour is fluctuation-dominated  and that the point free energy, unlike the one previously obtained within MFT, diverges in the limit of large chemical structure's width ($2L$) provided the chemical impurity is wetted by the liquid. The derived effective interaction potential is shown highly universal - for asymptotically large $L$ it does not depend on any of the system's parameters.

\section{Model}
\renewcommand{\theequation}{2.\arabic{equation}} 
\setcounter{equation}{0}
\vspace*{0.5cm}
The system under study consists of a two-dimensional fluid in a thermodynamic state infinitesimally away from its bulk liquid-vapour coexistence. The fluid occupies the half-space above a chemically inhomogeneous one-dimensional planar substrate composed of a chemically different domain of width $2L$ placed on an otherwise homogeneous substrate. The domain is referred to as type 1 substrate, while the remaining part of the solid substrate is of type 2. The substrate extends along the $x$-axis from $-X$ to $X$, where $X$ is assumed much larger than $L$, see Fig.1. One of the phases, say the liquid, is preferentially adsorbed at the substrate, while the gas phase remains stable in the bulk. We denote the wetting temperatures of (infinite) substrates 1 and 2 by $T_{W1}$ and $T_{W2}$, respectively, and assume $T_{W1}<T_{W2}$ while the fluid temperature $T$ is taken to be lower than $T_{W2}$.  

\begin{figure}[h!]
\begin{center}
\includegraphics[width=0.75\textwidth]{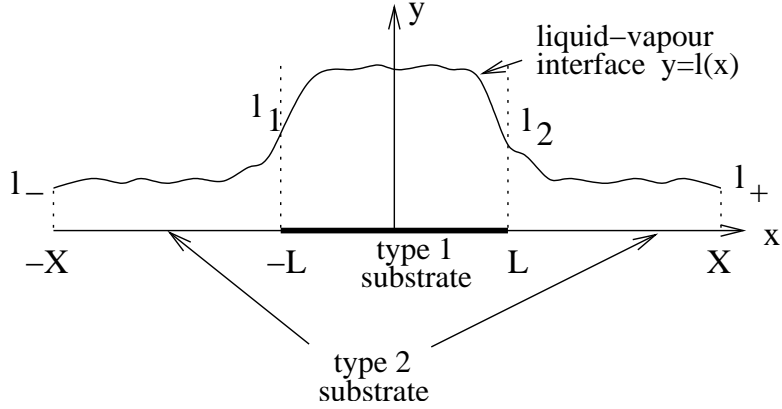}
\caption{A planar, chemically inhomogeneous substrate is exposed to fluid at its bulk liquid-vapour coexistence. A layer of a liquid-like phase is adsorbed near the substrate. The interface at $y=l(x)$ separates the adsorbed liquid-like layer from the gas phase which is stable in the bulk. The symbols $l_1$, $l_2$, $l_-$, $l_+$ denote the liquid layer height at $x=-L$, $+L$, $-X$ and $+X$, respectively.} 
\end{center}
\end{figure}
In this paper we consider the statistical mechanics of the liquid-vapour interface based on the SOS Hamiltonian
\beq
\label{Hamiltonian}
\mathcal{H}[l]=\int_{-X}^X dx\Big[\frac{\sigma}{2}\Big(\frac{dl}{dx}\Big)^2+V(x,l)\Big]\; ,
\eeq
where it is assumed that the system's state corresponds to a single-valued positive function $y=l(x)$ describing position of the liquid-vapour interface over the substrate. Here $\sigma$ is the interfacial stiffness parameter, and $V(x,l)$ denotes the interface's potential energy favoring its location near the wall. In the present model $V(x,l)$ takes - for $-X\leq x\leq X$ - the following form
\beq
\label{Potential}
 V(x,l)=\Theta (L-|x|)V_1(l)+\Theta (|x|-L) V_2(l)\; .  
\eeq
The corresponding partition function with the interface endpoints fixed at $(-X, l_-)$ and $(X, l_+)$ is given by the path integral
\beq
\label{Partition}
Z_0(l_-, l_+, X, L)= \int\mathcal{D}l e^{-\mathcal{H}[l]}\;,
\eeq
where the factor $\frac{1}{k_BT}$ is included into the Hamiltonian (\ref{Hamiltonian}). In the case of periodic boundary conditions $l_-=l_+$, and upon relaxing the constraint of fixed endpoint positions, the partition function can be written in the following form
\beq
\label{Partition2}
Z(X,L)=\int dl_- dl_1 dl_2 Z_2(l_-, l_1, X-L)Z_1 (l_1, l_2,2L)Z_2(l_2, l_-, X-L)\;,
\eeq
where $l_1=l(-L)$ and $l_2=l(L)$ are positions of the interface at $x=-L$ and $x=L$, respectively. The symbol $Z_i(y_1, y_2, \lambda)$ denotes partition function of an interface fluctuating above a homogeneous substrate of type $i=1,2$, of width $\lambda$ with its endpoints pinned at $y_1$ and $y_2$.  

Upon imposing the ''initial condition'' $Z_i(y_1, y_2,0)=\delta (y_1-y_2)$, each function $Z_i(y_1, y_2, \lambda)$ gains the following spectral representation 
\beq
\label{Spectral}
Z_i(y_1, y_2, \lambda)=\sum_n \psi_n^{(i)}(y_1)\psi_n^{(i)*}(y_2)e^{-E_n^{(i)}\lambda}
\eeq
in terms of orthonormal set of functions solving the Schr\"odinger equation 
\beq
\label{Shrodinger}
\Big[\frac{1}{2\sigma} \frac{d^2}{d y^2}-V_i(y)+E_n^{(i)}\Big]\psi_n^{(i)} (y)=0\;.
\eeq
In the present approach the potentials $V_i(y)$ have the following simple form
\beq
\label{Potential2}
V_i(y)=\left\{
\begin{array}{lll}
\infty\quad \textrm{for} \quad y<0  \\
-U_i/k_BT \quad \textrm{for} \quad 0<y<a_i \\
0 \quad \textrm{for} \quad a_i<y \;, 
\end{array}
\right.
\eeq
where $U_i$ and $a_i$ are positive constants. The above model was analyzed in detail by Burkhardt \cite{Burk} for the homogeneous substrate. In this case the partition function takes in the contact potential limit $a_i\to 0$, $U_i\to\infty$, $U_ia_i^2=const$, the following form   
\begin{eqnarray}
\label{BurkhardtZ}
Z_i(y_1,y_2,\lambda)=\sqrt{\frac{\sigma}{2\pi\lambda}}\left( e^{-\sigma (y_2-y_1)^2/2\lambda}+e^{-\sigma(y_1+y_2)^2/2\lambda}\right)+\nonumber\\
-\tau_ie^{\tau_i^2\lambda/2\sigma+\tau_i(y_2+y_1)}\textrm{erfc}\left[\sqrt{\frac{\sigma}{2\lambda}}(y_1+y_2)+\tau_i\sqrt{\frac{\lambda}{2\sigma}}\right]\;,
\end{eqnarray}
where erfc[x] stands for the complementary error function
\beq
\textrm{erfc[x]}=\frac{2}{\sqrt{\pi}}\int_x^\infty dte^{-t^2}
\eeq 
The parameter $\tau_i$ measures deviation of the system's temperature from the wetting temperature corresponding to substrate type $i$, i.e. $\tau_i\sim T-T_{Wi}$ (see Ref.\cite{Burk}). Within the present framework it is defined by
\beq
\label{Tau}
\left.\tau_i=\frac{1}{\psi_0^{(i)}}\frac{d\psi_0^{(i)}}{dy}\right|_{y=a_i}\;,
\eeq
where $\psi_0^{(i)}$ denotes the ground-state wave function for the $i$-th substrate. The considered range of temperatures corresponds to the situation in which only the ground states $\psi_0^{(i)}$ ($i=1,2$) may remain bound. In this case $\psi_0^{(i)}(y)=\sqrt{-2\tau_i}e^{\tau_i y}$ and the ground-state energy $E_0^{(i)}=-\frac{\tau_i^2}{2\sigma}$.  At the wetting temperature $T_{Wi}$ the bound state $\psi_0^{(i)}$  disappears in the continuum of the scattering solutions to Eq.(\ref{Shrodinger}) and $E_0^{(i)}\to 0^- $ - see \cite{Burk},\cite{deG}.   

We now insert the spectral representations (\ref{Spectral}) corresponding to \\
$Z_2(l_-, l_1, X-L)$ and $Z_2(l_2, l_-, X-L)$ into Eq.(\ref{Partition2}), and take advantage of the assumptions that $X\gg L$ and $T<T_{W2}$. Under these conditions the energy $E_0^{(2)}$ of the bound state $\psi_0^{(2)}$ is separated from the rest of the spectrum and the partition function $Z(X,L)$ takes the form
\beq
Z(X,L)=e^{-2E_0^{(2)}(X-L)}\int dl_1 \int dl_2\psi_0^{(2)*}(l_1)Z_1(l_1,l_2,2L)\psi_0^{(2)}(l_2)\; ,
\eeq
where terms of the order $e^{E_0^{(2)}(X-L)}$ have been neglected.
The excess free energy $\eta (L,T)$ due to chemical inhomogeneity is calculated by subtracting the line terms $2E_0^{(1)}L$ and $2E_0^{(2)}(X-L)$ from the system's free energy $F=-\ln Z(X,L)$. The subtracted terms correspond to energies of the homogeneous substrates of type $i=1,2$ of length $2L$ and $2(X-L)$, respectively, with the planar interfaces at their equilibrium positions.  Here $E_0^{(1)}=0$ for $\tau_1\geq 0$.

This way one obtains the following formula for $\eta (L,T)$.
\beq
\label{eta000}
\eta (L,T) =-2E_0^{(1)}L-\Bigg[\ln\int_0^\infty\int_0^\infty dl_1 dl_2\psi_0^{(2)*}(l_1)Z_1(l_1,l_2,2L)\psi_0^{(2)}(l_2)\Bigg]\;,
\eeq
which will be analyzed in the following section. 

\section{Excess point free energy}
\renewcommand{\theequation}{3.\arabic{equation}} 
\setcounter{equation}{0}
\vspace*{0.5cm}

The integral in Eq.(\ref{eta000}) can be evaluated by substituting the expression (2.8) for $Z_1(l_1,l_2,2L)$, changing the integration variables $l_1=\frac{u-v}{2}$, $l_2=\frac{u+v}{2}$ and then integrating by parts. As a result one obtains the following formula
\beq
\label{etaOG}
\eta= -2E_0^{(1)}L-\ln\Bigg\{\frac{1}{\sigma (\tau_1+\tau_2)^2}\Big[2\tau_2^2(\tau_1^2-\tau_2^2)L+\sigma (\tau_1^2+\tau_2^2)\Big]e^{\frac{2L}{\xi_{||2}}}\textrm{erfc}\Big[\sqrt{\frac{2L}{\xi_{||2}}}\Big]+
\eeq
$$
-\frac{\tau_1-\tau_2}{\tau_1+\tau_2}\sqrt{\frac{2L}{\xi_{||2}}}+\frac{2\tau_1\tau_2}{(\tau_1+\tau_2)^2}e^{\tau_1^2\frac{L}{\sigma}}\textrm{erfc}\Big[\tau_1\sqrt{\frac{L}{\sigma}}\Big]\Bigg\}\;,
$$
in which the parameter $\frac{2\sigma}{\tau_2^2}$ has been identified as the parallel correlation length $\xi_{||2}$ corresponding to interfacial fluctuations above homogeneous substrate type 2 \cite{Burk}. From Eq.(\ref{etaOG}) it follows, that at fixed temperature the quantity $\eta (L,T)$ is a non-negative, increasing, and concave function of the width $2L$. Whether $\eta$ is bounded from above depends on the sign of $\tau_1$. This issue is discussed in the following subsection.

\subsection{Large $L$ regime}

In the asymptotic regime of large $L$ the behaviour of the quantity $\eta$ given by (\ref{etaOG}) crucially depends on the sign of $\tau_1$. This is easily understood by noting that the argument of the error function diverges either to plus or minus infinity depending on the sign of $\tau_1$. Also the value of the ground state energy $E_0^{(1)} $ is negative and equal $-\frac{\tau_1^2}{2\sigma}$ for $\tau_1<0$, and equal zero in the opposite case. To obtain the asymptotic expression for $\eta$, one expands the formula (\ref{etaOG}) for $L$ such that $\frac{L}{\xi_{||2}}\gg 1$ and $\frac{L\tau_1^2}{\sigma}\gg 1$ (the case $\tau_1=0$ is analyzed separately). This way we obtain the following expressions:
\beq
\label{eta}
\eta=
\left\{
\begin{array}{ll}
\ln\frac{(\tau_1+\tau_2)^2}{4\tau_1\tau_2}+\frac{1}{4\sqrt{\pi}}\frac{(\tau_1+\tau_2)^2(\tau_1-\tau_2)^2}{\tau_1^3\tau_2^4}\Big(\frac{\sigma}{L}\Big)^{3/2}e^{-2L/\xi_{||1}}\quad \textrm{\hspace{0.4cm}for}\quad \tau_1<0\\
\frac{1}{2}\ln \Big(\frac{\tau_2^2\pi}{4\sigma}L\Big)\quad \textrm{\hspace{6.2cm}for}\quad \tau_1=0 \\
\ln \Big(-\sqrt{\pi}\frac{\tau_1^2\tau_2^3}{(\tau_1-\tau_2)^2}\Big(\frac{L}{\sigma}\Big)^{3/2}\Big)\quad \textrm{\hspace{3.6cm} for}\quad \tau_1>0\;,
\end{array}
\right. 
\eeq
where $\xi_{||1}=\frac{2\sigma}{\tau_1^2}$ provided $\tau_1<0$. The central observation regarding the above formula is the change of the character of the $L$-dependent contribution to the quantity $\eta$. In the temperature regime $T<T_{W1}$ it converges to a finite and positive value $2\eta_0= \ln\frac{(\tau_1+\tau_2)^2}{4\tau_1\tau_2}$, which is a symmetric function of $\tau_1$ and $\tau_2$. Actually $\eta_0$ can be considered as the point tension due to a single inhomogeneity at which two semi-infinite, homogeneous substrates of type 1 and 2 meet \cite{K1}, \cite{D2}.

 The finite $L$ correction is negative and decays as $(\frac{1}{L})^{3/2}e^{-2L/\xi_{||1}}$, where $\xi_{||1}$ sets the length scale controlling the rate of the decay. On the other hand, in cases $\tau_1=0$ and $\tau_1>0$, in which $\xi_{||1}$ becomes infinite, one finds the quantity $\eta$ to be logarithmically divergent in the limit $L\to\infty$. This divergence is attributed to fluctuations, as it is absent within the mean-field approach which will be discussed in Section 4. Let us also note that the derivative $-\frac{\partial\eta}{\partial 2L}$ has the interpretation of an effective force acting between two interface heterogeneities concentrated around $x=-L$ and $x=L$. This force is attractive for all of the considered temperature regimes. It decays as $(\frac{1}{L})^{3/2}e^{-2L/\xi_{||1}}$ for $T<T_{W1}$, while in the other cases the decay is of the type $\frac{1}{L}$. This force has universal properties - it depends neither on the substrates', nor fluid's characteristics. More explicitly, it equals $-\frac{1}{2L}$ for $T=T_{W1}$ and $-\frac{3}{2L}$ in the case $T>T_{W1}$ (in the units where $k_BT=1$).   

\subsection{Small $L$ regime}
Within the present framework it is also possible to analyze the excess point free energy in the regime of system's parameters corresponding to the inhomogeneity's width $2L$ small compared to the lengthscales $\frac{\sigma}{\tau_1^2}$, $\frac{\sigma}{\tau_2^2}$. The regime of small $L$ in the case $\tau_1=0$ is explored separately. 
The coefficient $\eta$ vanishes in the limit $L\to 0$, as expected on physical grounds. The formula (3.1) in the case of asymptotically small $L$, has the following form
\beq
\eta=\frac{(\tau_1-\tau_2)^2}{\sigma}L+...\; ,
\eeq
where terms of higher order have been omitted. It follows, that $\eta$ decays linearly to zero as $L\to 0^+$ and the corresponding coefficient has the same form in different temperature regimes.   

\section{Mean field theory}
\renewcommand{\theequation}{4.\arabic{equation}} 
\setcounter{equation}{0}
\vspace*{0.5cm}
In this section we compare the conclusions of the previous section with results for the quantity $\eta$ obtained within mean-field approximation based on the interfacial Hamiltonian
\beq
\label{MFTHam}
 \mathcal{H}[l]=\int dx 
\, \Bigg[\frac{\Sigma}{2}\Big(\frac{dl}{dx}\Big)^2+\omega(x,l)\Bigg] \quad , 
\eeq
where $\Sigma$ is the liquid-vapour surface tension, and $\omega (x,l)$ denotes the effective potential of interaction between the substrate and the liquid-vapour interface. It is approximated by a piecewise constant function of $x$ \cite{D2},\cite{P1}
\beq
\label{omega}
\omega(x,l)=\Theta (L-|x|)\omega_1(l)+\Theta (|x|-L) \omega_2(l),
\eeq
where - in the case of short-ranged intermolecular interactions - $\omega_i$ $(i=1,2)$ has the following structure:
\beq
\label{omegai}
\omega_i(l)= \tau_i e^{-l/\xi_B}+b e^{-2 l/\xi_B}+... \quad .
\eeq
Here $\xi_B$ denotes the bulk correlation 
length in the adsorbed liquid phase and the positive parameter $b$ takes account of 
repulsion of the interface from the substrate at short distances. 
For simplicity, $b$ is assumed to have the same value for both substrate types. The mean-field analysis of Eq.(\ref{MFTHam}), see Refs \cite{P1},\cite{P2}, leads to the following conclusions:\\
- The mean-field expression for $\eta$ converges in the limit $L\to\infty$ to a finite positive function $2\eta_0$, where the point tension $\eta_0$ depends on the  model parameters $\tau_1$, $\tau_2$, $\Sigma$, $b$.\\
- The dominant $L$-dependent correction to $2\eta_0$ is negative. For temperatures $T<T_{W1}$ it is of the order $e^{-2L/\xi_{||1MFT}}$ with amplitude depending on all system's parameters; here $\xi_{||1MFT}=-\frac{\sqrt{2\Sigma b}}{\tau_1}$. In the remaining temperature regimes, i.e. for $T_{W1}\leq T < T_{W2}$, the corrections are of the order $\frac{1}{L}$, and their magnitude is governed by a universal prefactor $\Sigma\xi_B^2$. 

Comparing these predictions to formula (\ref{eta}) one concludes, that the singular behaviour of $\eta$ occurring for $\tau_1\geq 0 $ and $L\to\infty$ is due to fluctuations. It is also worth noting that fluctuations wash out the dependence of $\eta$ on the interfacial stiffness. Indeed - the energy scale $\Sigma\xi_B^2 $ controlling the $L$-dependent terms in $\eta$ within the mean-field  description is replaced by $k_B T$ (which was put equal to unity in the calculation leading to Eq.(\ref{eta})). As regards the temperature regime $\tau_1<0$ - one observes the presence of the prefactor $(\frac{1}{L})^{3/2}$ in the expression for $\eta$, which is absent at mean-field level. The length controlling the exponential decay of the $L$-dependent terms in $\eta$ is in both cases set by the correlation length describing the range of typical fluctuations of the interface over homogeneous substrate type 1. 

\section{Remarks on mean-field results for $\tau_1<0$}
\renewcommand{\theequation}{5.\arabic{equation}} 
\setcounter{equation}{0}
\vspace*{0.5cm}
In this section we introduce a simplified interfacial model, within which we reproduce results qualitatively equivalent to the mean-field predictions in the case $\tau_1<0$. Physically the procedure proposed below amounts to damping the long-ranged fluctuations by an external potential $W(x,l)$, but still summing all the relevant terms in the partition function. Within mean-field approach one does something else, namely takes only the dominant term from the partition trace. These complementary approaches yield equivalent results regarding the dependence of $\eta$ on $L$.

Although the calculation presented in this section is limited to the temperature regime $T<T_{W1}$, the considerations are on general grounds as no explicit form of the interfacial potential is applied.
We consider a liquid-vapour interface fluctuating in presence of an effective potential $W(x,l)$, which is assumed to have the piecewise-constant structure:
\beq   
W(x,l)=\Theta (L-|x|)W_1(l)+\Theta (|x|-L) W_2(l)\;.
\eeq
The potentials $W_i(l)$ ($i=1,2$) have single, well localized  minima at $l_{\pi i}$, whose role is to damp the fluctuations and to localize the average interface's position at $l_{\pi 1}$ and $l_{\pi 2}$, respectively, for $|x|<L$ and $|x|>L$ . The system's Hamiltonian is given by (\ref{Hamiltonian}) with $V(x,l)$ substituted by $W(x,l)$. The  partition function is written in the form of Eq.(\ref{Partition2}) and the spectral decompositions of $Z_1$, $Z_2$ are inserted into it. The key characteristic of the potentials $W_i(l)$ is that the corresponding Shr\"odinger operators have only discrete spectra. It follows, that the partition function takes the following form
\begin{displaymath}
Z=e^{-2E_0^{(2)}(X-L)}\Bigg(1+\mathcal{O}\Big(e^{(2E_0^{(2)}-2E_1^{(2)})(X-L)}\Big)\Bigg)\int dl_1\int dl_2\Big[\phi_0^{(2)}(l_1)\phi_0^{(2)*}(l_2)\Big]\times
\end{displaymath}
\beq
\label{Zet}
\times \Big[\phi_0^{(1)}(l_2)\phi_0^{(1)*}(l_1)e^{-2E_0^{(1)}L}+\phi_1^{(1)}(l_2)\phi_1^{(1)*}(l_1)e^{-2E_1^{(1)}L}+\mathcal{O}\Big(e^{-2E_2^{(1)}L}\Big)\Big]\;,
\eeq  
where the eigenfunctions $\phi_j^{(i)}$ and the eigenvalues $E_j^{(i)}$ are computed from the corresponding Shr\"odinger equation
\beq
\Big[\frac{1}{2\sigma} \frac{d^2}{d y^2}-W_i(y)+E_j^{(i)}\Big]\phi_j^{(i)} (y)=0\;.
\eeq 
It is then clear that the quantity $\eta$ 
\beq
\eta=\lim_{X\to\infty}[-\ln Z-2E_0^{(1)}L-2E_0^{(2)}(X-L)]
\eeq
converges to 
\beq
2\eta_0=\int dl_1\int dl_2\phi_0^{(2)}(l_1)\phi_0^{(2)*}(l_2)\phi_0^{(1)}(l_2)\phi_0^{(1)*}(l_1) < \infty
\eeq
for $L\to\infty$ and that the dominant $L$-dependent contribution to $\eta$ is given by
\beq
e^{-2|E_0^{(1)}-E_1^{(1)}|L}\int dl_1\int dl_2\phi_0^{(2)}(l_1)\phi_0^{(2)*}(l_2)\phi_1^{(1)}(l_2)\phi_1^{(1)*}(l_1)\;.
\eeq
The energy gap $E_1^{(1)}-E_0^{(1)}>0$ corresponds to inverse of the interfacial correlation length $\xi_{||1}$ \cite{Burk}. All the next-to-leading finite $L$ corrections to $\eta$ are also easily computed from (\ref{Zet}). The magnitude of the $j$-th term is of the order $e^{-2L|E_j^{(1)}-E_0^{(1)}|}$ for $j=2,3,...$ . 

The above calculation shows that the general characteristics of the $L$-dependent contributions to $\eta(L,T)$, namely their exponential decay controlled by $\xi_{||1}$ do not depend on the explicit form of the short-ranged interfacial potential. The only assumptions made above are the piecewise-constant form of $W(x,l)$ in Eq.(5.1) and discrete spectra of the Schr\"odinger operators corresponding to $W_{i}(l)$. The later implies that $\lim_{l\to\infty}W_i(l)=\infty$, which results in damping the long-ranged fluctuations. As a consequence even with fluctuations one recovers the mean-field theory result rather than the one obtained in Section 3 for the potential in Eq.(2.7) which has both discrete and continuous spectrum. These two results differ by the factor $L^{-3/2}$, which is due to the continuous spectrum contribution.

\section{Contact point height}
\renewcommand{\theequation}{6.\arabic{equation}} 
\setcounter{equation}{0}
\vspace*{0.5cm}
A remarkable feature of the mean-field interfacial height $\bar{l}(x=L)$ at the contact point where the substrates of type 1 and 2 meet is that it remains finite for all $L$ values and all temperatures. On the other hand, the interfacial height $\bar{l}(x=0)$ at the domain's center diverges logarithmically in the limit $L\to\infty$ provided $T\geq T_{W1}$ \cite{P1}. Whether these conclusions remain valid in the considered two-dimensional case, where fluctuations play an important role, has to be checked. In the following, within the SOS model introduced in Section 2, we construct the probability density $P_L(l)$ of finding the interface at height $l$ for $x=L$. Once $P_L(l)$ is known, the average interface position as well as higher moments can be expressed in terms of $L$ and $T$.   

The probability density corresponding to $x=\pm L$ is given by
\beq
P_L(l)=\frac{\int\mathcal{D} le^{-\mathcal{H}[l]}\delta (l(L)-l)}{\int \mathcal{D}l e^{-\mathcal{H}[l]}}\;,
\eeq
which for the case $\tau_2<0$ and upon neglecting terms that vanish in the limit $X\to\infty$, can be written as 
\beq
P_L(l)=\frac{\int dl_1\int dl_2 \psi_0^{(2)}(l_1)\psi_0^{(2)*}(l_2)Z_1(l_2,l_1,2L)\delta(l_2-l)}{\int dl_1\int dl_2\psi_0^{(2)}(l_1)\psi_0^{(2)*}(l_2)Z_1(l_2,l_1,2L)} \;.
\eeq
Inserting the asymptotic form of $Z_1(l_2,l_1,2L)$
\begin{displaymath}
Z_1(l_2,l_1,2L)=-2\tau_1\Theta(-\tau_1)e^{\tau_1^2L/\sigma}e^{\tau_1(l_1+l_2)}+
\end{displaymath}
\beq
+\frac{1}{2\sqrt{\pi}}\Big(\frac{\sigma}{L}\Big)^{3/2}(l_1+\tau_1^{-1})(l_2+\tau_1^{-1})+\mathcal{O}(L^{-5/2})
\eeq
and performing the integration yields the following formula
\beq
P_L(l)=-\frac{[\frac{4\tau_1\tau_2}{\tau_1+\tau_2}\Theta(-\tau_1)e^{\tau_1^2L/\sigma}e^{\tau_1 l}+\frac{1}{\sqrt{\pi}}(\frac{\sigma}{L})^{3/2}(l+\tau_1^{-1})(\tau_2^{-1}-\tau_1^{-1})]e^{\tau_2 l}}{\frac{4\tau_1\tau_2}{(\tau_1+\tau_2)^2}\Theta(-\tau_1)e^{\tau_1^2L/\sigma}-\frac{1}{\sqrt{\pi}}(\frac{\sigma}{L})^{3/2}\tau_2^{-1}(\tau_2^{-1}-\tau_1^{-1})^2}\;.
\eeq
 The average position of the interface at $x=\pm L$ is given by 
\beq
<l>_L=
\left\{
\begin{array}{ll}
-\frac{1}{\tau_1+\tau_2}(1+\mathcal{O}(e^{-L/\xi_{||1}}))\quad \textrm{for} \quad \tau_1<0 \\
-\frac{2\tau_1-\tau_2}{\tau_2(\tau_1-\tau_2)}(1+\mathcal{O}(1/L)) \quad \textrm{for} \quad \tau_1>0\;.
\end{array}
\right.
\eeq
and the standard deviation $A(L,T)=\sqrt{<l^2>_L-<l>_L^2}$ converges to a finite value in the limit $L\to\infty$ in both cases. Interestingly, for $\tau_1<0$ one obtains 
\beq
\lim_{L\to\infty}A(L,T)=\lim_{L\to\infty}<l>_L\;.
\eeq
From the above results one concludes that no sign of critical divergences is present in the contact regions corresponding to $x\simeq \pm L$. Both the height $<l>_L$ and the fluctuations' magnitude remain finite in the limit $L\to\infty$ at all temperatures. It follows that for non-negative $\tau_1$ the behaviour of the coefficient $\eta$ is strongly influenced by the system's critical fluctuations that occur only far away from the contact region, to which the line or point tensions are usually attributed.

\section{Summary}
\renewcommand{\theequation}{7.\arabic{equation}} 
\setcounter{equation}{0}
\vspace*{0.5cm} 
In this paper we were concerned with properties of the excess point free energy $\eta$ for adsorption at a one-dimensional substrate equipped with a single stripe-like chemical inhomogeneity. Our conclusions are restricted to short-ranged intermolecular interactions and to thermodynamic states at the liquid-vapour coexistence. Under these conditions our predictions are as follows.
\begin{itemize}
\item
At fixed temperature the excess point free energy is a positive, increasing, concave function of the heterogeneity's width $2L$. It converges to a finite value in the limit $L\to\infty$ provided the stripe remains non-wetted, i.e. $T<T_{W1}$. The dominant $L$-dependent correction is negative and decays as $L^{-3/2}e^{-2L/\xi_{||1}}$, where the decay rate is controlled by the interfacial correlation length corresponding to type 1 substrate (i.e. the stripe).  

\item 
For temperatures $T_{W1}\leq T <T_{W2}$  the excess point free energy exhibits a logarithmic singularity in the limit of macroscopic domain sizes, i.e. $L\to\infty$. This singularity is attributed to fluctuations and it is absent within mean-field approach. 

\item 
The derivative $-\frac{\partial\eta}{\partial 2L}$ is analogous to an effective solvation force acting between the adsorbed fluid's heterogeneities at $x\simeq -L$ and $x\simeq L$. This force is attractive in all temperature regimes. It decays as $L^{-3/2}e^{-2L/\xi_{||1}}$ in the case $T<T_{W1}$, while for $T\geq T_{W1}$ it is proportional to $1/L$, with universal amplitudes (i.e. amplitudes independent of the substrates' and fluid's properties). 

\item
The asymptotic decay of excess point free energy in the regime of heterogeneity's width small compared to parallel correlation lengths present in the system is linear in $L$ for all temperature regimes. 

\item
The results are compared to mean-field predictions. The basic observation  is occurrence of the fluctuation-induced divergence of $\eta$ in the limit $L\to\infty$ for $T_{W1}\leq T <T_{W2}$.
 Moreover, incorporating fluctuations washes out the dependence of the force's amplitude on the adsorbed fluid's properties described by the interfacial tension $\Sigma$ and the bulk correlation length $\xi_B$. This means that the solvation force asymptotically does not depend on any of the system's parameters at all. On the other hand, both with and without fluctuations one observes crossover of the force from exponential to algebraic decay at $T=T_{W1}$. 

\item
We proposed a phenomenological  way of obtaining the mean-field results in the temperature regime $T<T_{W1}$. It amounts to trapping the fluctuating interface in a single minimum potential $W(x,l)$. As a result the law describing the decay of the dominant $L$-dependent contribution to $\eta$ turns out to be the same for many reasonable choices of $W(x,l)$.

\item
In order to characterize the system's state in the vicinity of the substrate inhomogeneities $x=\pm L$ we calculated the probability distribution function $P_L(l)$ for large $L$ and showed that the average separation $<l>_L$ of the interface from the substrate, as well as the fluctuations magnitude $<l^2>_L-<l>_L^2$, remain bounded in the limit $L\to\infty$ for all temperatures $T<T_{W2}$. This means that for $T\geq T_{W1}$ the transition regions $x\simeq\pm L$ contain no indication of the critical state of the interface around $x=0$. This result is consistent with mean-field predictions. 
    
\end{itemize}

Although a direct link to experimental studies performed so far (e.g. Refs [3-6]) does not seem possible at this stage, we believe that some of our predictions are open to experimental verification. For this purpose let us point that the results may be referred to a system in which the solid substrate is substituted with a non-wettable Langmuir-Blodgett film decorated with a wettable stripe whose compressibility is orders of magnitude larger than in the case of a solid structure. The adsorbed wetting layer will tend to shrink the inhomogeneity due to the effective force acting between the linear heterogeneities around $x\simeq -L$ and $x\simeq L$. This effect should become pronounced once the temperature $T_{W1}$ is exceeded. Our predictions regarding the properties of the force $-\frac{\partial\eta}{\partial (2L)}$ are in principle open to verification via direct measurement of the stripe compression. Certainly a three-dimensional system is experimentally more accessible and the influence of fluctuations on the quantity $\eta$ is not clear in this case. On the other hand both mean-field theory and the exact calculation in two dimensions indicate appearance of the long-ranged effective force of universal properties. The same kind of behaviour can be expected in a physical system in three dimensions.\\

\noindent {\bf {Acknowledgment}}  \\ 
\noindent This work has been supported by the  Ministry of Science and Higher
Education via the grant N202 076 31/0108.

\end{document}